\documentclass[a4paper,11pt]{article}

\usepackage{pos}
\usepackage{graphicx}
\usepackage{subcaption}
\usepackage{lineno}
\usepackage{siunitx}
\usepackage[capitalise]{cleveref}
\usepackage{lineno}

\newcommand{\lh}{\mathcal{L}}
\newcommand{\nunu}{$\nu_{\alpha}\bar{\nu}_{\alpha}$}
\newcommand{\bb}{$b\bar{b}$}
\newcommand{\ww}{$W^{-}W^{+}$}
\newcommand{\tautau}{$\tau^{-}\tau^{+}$}
\newcommand{\numu}{$\nu_{\mu}$}

\newcommand{\charon}{\texttt{\raisebox{0.82\depth}{$\chi$}aro$\nu$}}
\newcommand{\nusquids}{\texttt{nuSQuIDS}}

\title{Limits on WIMP-Scattering Cross Sections using Solar Neutrinos with Ten Years of IceCube Data}
\ShortTitle{IceCube Solar DM}

\author{The IceCube Collaboration \\{\normalsize \normalfont(a complete list of authors can be found at the end of the proceedings)}\\}

\emailAdd{jeffrey.lazar@uclouvain.be}

\abstract{Although dark matter (DM) comprises 84\% of the matter content of the Universe, its nature remains unknown.
One broad class of particle DM motivated by extensions of the Standard Model (SM) is weakly interacting massive particles (WIMPs).
Generically, WIMPs will scatter off nuclei in large celestial bodies such as the Sun, thus becoming gravitationally bound. 
Subsequently, WIMPs can annihilate to stable SM particles, ultimately releasing most of their energy as high-energy neutrinos which escape from the Sun.
Thus, an excess of neutrinos from the Sun's direction would be evidence for WIMPs.
The IceCube Neutrino Observatory is well-suited to such searches since it is sensitive to WIMPs with masses in the region preferred by supersymmetric extensions of the SM.
I will present the results of IceCube's most recent solar WIMP search, which includes all neutrino flavors, covers the WIMP mass range from \SI{20}{\GeV} to \SI{10}{\TeV}, and has world-leading sensitivity over this entire range for most channels considered.

\vspace{4mm}

{\bfseries Corresponding authors:}
Jeffrey Lazar$^{1*}$\\
{$^{1}$ \itshape Université Catholique de Louvain, Pl. de l'Université 1, 1348 Ottignies-Louvain-la-Neuve}\\[4mm]
$^*$ Presenter
}

\ConferenceLogo{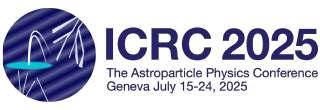}

\FullConference{39th International Cosmic Ray Conference (ICRC2025)\\
 15–24 July 2025\\
Geneva, Switzerland\\}

\begin{document}
\maketitle

\section{Introduction}
\label{sec:introduction}

Dark matter (DM) makes up about 84\% of all matter in the Universe~\cite{Freese:2017idy} and may consist of fundamental particles~\cite{Bertone:2004pz}. 
A worldwide effort is underway to detect these particles~\cite{Liu:2017drf, Conrad:2017pms,Buchmueller:2017qhf,Arguelles:2019ouk}, with particular interest in weakly interacting massive particles, which interact with Standard Model (SM) particles at or below the weak scale.
If DM is comprised of such particles, they could scatter off nuclei in the Sun, lose energy, and become gravitationally trapped by its gravity~\cite{Jungman:1995df}.
Through continued interactions, DM would accumulate at the Sun’s core.
Eventually, these particles would annihilate into SM particles, but only neutrinos, due to their weak interactions, could escape the dense solar interior. 
This opens a pathway for indirect DM detection by observing an excess of high-energy neutrinos coming from the Sun.
Because DM capture and annihilation rates in the Sun are expected to reach equilibrium~\cite{Jungman:1995df}, such neutrino-based searches are sensitive to the DM-proton scattering cross section.
This approach complements direct detection methods on Earth, which look for energy from nuclear recoils caused by DM interactions~\cite{Liu:2017drf}.

The IceCube Neutrino Observatory, located near the geographic South Pole, is a gigaton-scale Cherenkov detector embedded in the Antarctic ice.
It comprises 5,160 digital optical modules (DOMs) arranged along 86 vertical strings, each holding 60 modules.
These DOMs detect Cherenkov light emitted by relativistic charged particles produced in neutrino interactions.
The DOMs are spaced \SI{17}{\m} apart vertically in the main array on a hexagonal grid with \SI{125}{\m} between strings.
The denser DeepCore sub-array features a tighter configuration, with $\SI{7}{\m}$ vertical spacing and about $\SI{70}{\m}$ spacing between strings.
Together, the main array and DeepCore allow IceCube to detect neutrinos across a wide energy range, from \SI{5}{\GeV} to \si{PeV}.
This capability enables IceCube to probe a dark matter mass range comparable to that of direct detection experiments.

\section{Simulation and Data Sample}
\label{sec:simulation}

This analysis combines IceCube’s low-energy event selection---initially developed for measuring atmospheric neutrino oscillation parameters~\cite{IceCubeCollaboration:2024ssx}---with the high-energy selection used in searches for neutrino emission from astrophysical point sources~\cite{IceCube:2022der}.
The only modification to these selections is the addition of a filter to remove any overlap between the two selections, ensuring statistical independence.
The high-energy selection targets a high-purity sample of charged-current \numu{} events.
At energies above $\approx\SI{300}{GeV}$, the resulting $\mu^{\pm}$ can travel more than a kilometer~\cite{Koehne:2013gpa}, creating long, track-like signals that allow for precise directional reconstruction.

\begin{figure}
    \centering

    \begin{subfigure}[t]{0.6\textwidth}
        \centering
        \includegraphics[
            width=\textwidth,
            height=\textheight,
            keepaspectratio]{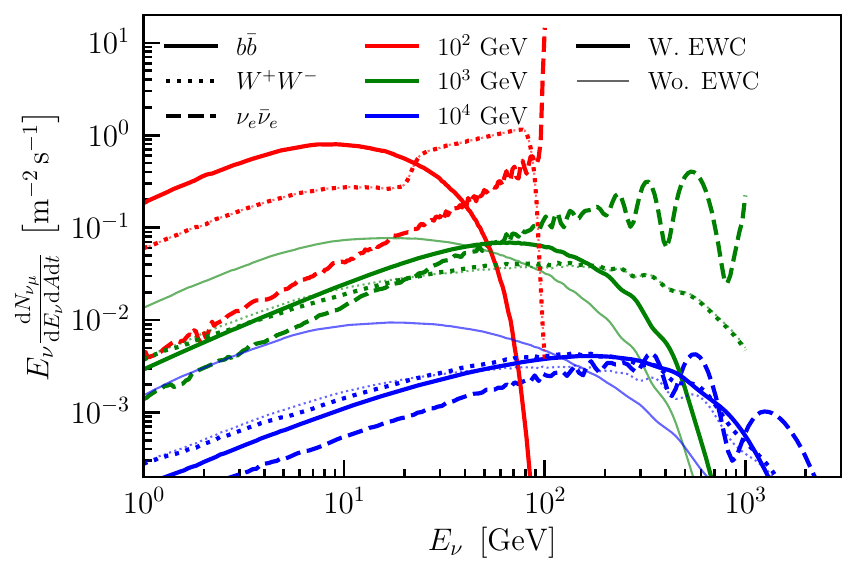}
    \end{subfigure}%
    \hfill
    \begin{subfigure}[t]{0.4\textwidth}
        \centering
        \includegraphics[
            width=\textwidth,
            height=\textheight,
            keepaspectratio]{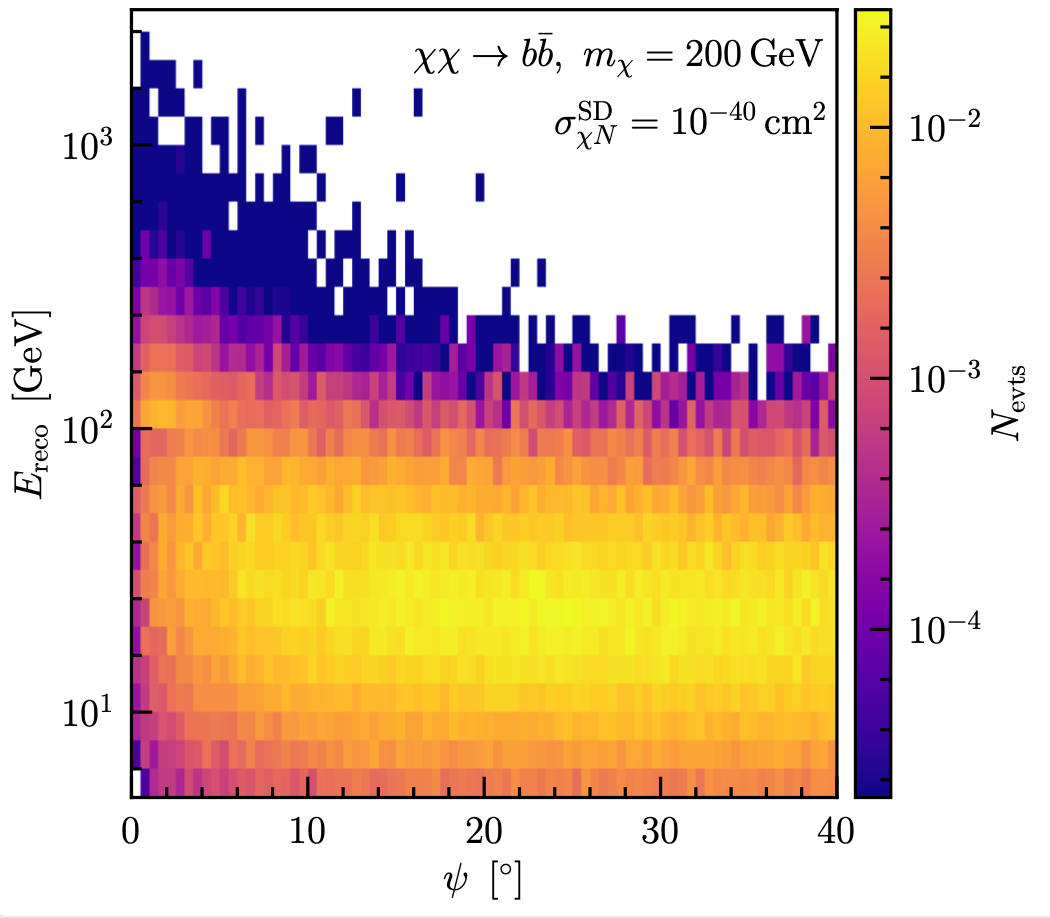}
    \end{subfigure}%
    
    \caption{\textbf{\textit{Spectra of neutrinos for several annihilation channels and WIMP masses at Earth and expected reconstructed event distribution.}}
    Left: The line styles represent a primary annihilation channel, while the colors represent different WIMP masses.
    Annihilation channels studied in previous IceCube analyses include a lower opacity line to show the spectra without the EW effect.
    Note the significant differences in the spectra from annihilation to \bb{} at masses above the EW scale and the absence of such differences from annihilation to \ww{}.
    Right: The expected distribution of events from 200 GeV DM annihilating to $b\bar{b}$ as a function of the reconstructed energy and reconstructed angular distance from the Sun.
    }
    \label{fig:signals}
\end{figure}

In contrast, the low-energy selection targets events in a regime where muons produce less light and travel shorter distances, reducing angular resolution.
As a result, this sample includes neutrinos of all flavors rather than focusing on \numu{} alone.

Within the analysis energy range of \SI{5}{\GeV} to \SI{10}{\TeV}, the most common background comes from atmospheric neutrinos and muons produced by cosmic-ray interactions in Earth's atmosphere~\cite{Gaisser:2002jj}.
These backgrounds are modeled using real data, with reconstructed right ascension randomized for each event.
This procedure is equivalent to time-scrambling the events in IceCube's local coordinate system.
Since the Sun lies far from the equatorial poles, this scrambling diminishes any solar clustering, effectively removing point-source signatures while preserving the declination dependence of background rates.
This approach minimizes systematic uncertainties related to background modeling.

In addition to these backgrounds, this analysis also includes the background from solar atmospheric neutrinos, which are produced by cosmic-ray interactions within the solar atmosphere.
To model this, this analysis adopts a recent model of the solar atmospheric neutrino flux~\cite{Arguelles:2017eao}, based on a hybrid solar density profile from Refs.~\cite{spruit:1974abc,Bahcall:2004pz,fontenla:2007abc,gingerich:1971abc} and computed using the \texttt{MCEq} cascade equation solver~\cite{Fedynitch:2015zma}.
Neutrinos are generated from interactions of primary hadrons and secondary muons and are then propagated to Earth using the density matrix formalism implemented in the \nusquids{}~\cite{Arguelles:2021twb} package.
The same framework handles the final propagation step from Earth’s surface to the detector.
The flux calculation also incorporates variations from cosmic-ray spectra and hadronic interaction models.
For this analysis, we use the flux resulting from the combined Gaisser-Honda with Hillas-Gaisser H4a primary cosmic-ray flux~\cite{Fedynitch_2012} and Sibyll-2.3c hadronic interaction model~\cite{Riehn:2015aqb,Riehn:2015oba} assuming a purely protonic atmosphere and MRS prompt model~\cite{Martin:2003us}.

To model the signal neutrinos from DM annihilation, this analysis employs the \charon{} package~\cite{Liu:2020ckq}.
This package assumes DM particles annihilate into a pair of monochromatic Standard Model particles.
The spectra of resulting particles are drawn from precomputed tables, and all unstable particles are allowed to interact and decay until only neutrinos remain.
This process uses a Monte Carlo method to model interactions and decays within the Sun’s dense core.
The resulting neutrino flux is then propagated to the detector using \nusquids{}.

\charon{} offers two approaches for generating the initial particle spectra: one using the PYTHIA 8.2 library~\cite{Sjostrand:2014zea} and another incorporating electroweak (EW) corrections~\cite{Bauer:2020jay}.
EW effects become significant for DM masses above the EW scale, where decays may involve SM particles before electroweak symmetry breaking.
While earlier models often neglected~\cite{Niblaeus:2019gjk} or only approximated these effects~\cite{Cirelli:2010xx}, \charon{} fully incorporates them.
This effect can substantially alter the predicted neutrino spectra for specific annihilation channels.
The left panel of \cref{fig:signals} compares the propagated spectra with and without these effects for three DM masses and annihilation channels.
At 100 GeV (below the EW scale), the spectra match across all channels. However, for 1 TeV and 10 TeV masses, a notable enhancement appears for the $b\bar{b}$ channel due to prompt weak boson emission producing hard neutrinos.
In the $W^{-}W^{+}$ channel, the spectra shift slightly as energy is distributed across more weak bosons, leading to a softer neutrino output.

\section{Analysis Methodology}
\label{sec:analysis_methods}

For the analysis, data are binned based on reconstructed quantities to build expected distributions for background, solar atmospheric neutrinos, and neutrinos resulting from dark matter annihilation.
In the high-energy selection, events are binned by reconstructed energy and their reconstructed angular separation from the Sun’s center.
An additional binning is applied for the low-energy selection using a morphological classification variable designed to distinguish charged-current \numu{} interactions from other event types.
This variable is particularly useful because neutrinos originating from the Sun arrive in a flavor-mixed state, whereas atmospheric neutrinos above $\approx\SI{40}{\GeV}$ tend to retain a \numu{} dominance.
Consequently, detecting an excess of non-\numu{} events at specific energies can help differentiate a potential signal from the background.

Distributions for solar atmospheric neutrinos and DM-induced neutrinos are derived from Monte Carlo simulations.
For each simulated event, a random time within the analysis window is selected, the Sun’s position is calculated, and only events occurring within the Sun’s physical extent are retained.
This process is repeated numerous times, and the average is taken to find the expected time-averaged distributions.

In addition to modeling the solar atmospheric flux, simulations are performed for 51 dark matter scenarios, with DM masses ranging from \SI{20}{\GeV} and \SI{10}{\TeV} and annihilation into \bb{}, \ww{}, \tautau{}, and \nunu{} channels.
The right panel of \cref{fig:signals} shows the predicted event rate for a benchmark DM model with a reference cross section of $\sigma_{\chi p}=10^{-40}\,\si{\cm\squared}$ for both event selections.
The low-energy distribution has been summed over the morphological discriminator for clarity in this figure.
At energies above \SI{100}{\GeV}, the high-energy selection dominates, as indicated by event clustering within \SI{3}{\degree} of the Sun.
At lower energies, where angular resolution decreases, events are distributed over a broader angular region.
A slight discontinuity at \SI{100}{\GeV} reflects a selection cut applied to the reconstructed energy in the high-energy sample.

The data-derived background distribution is computed by taking all events that pass the selections and assigning a random right ascension drawn uniformly from the distribution $\left[0, 2\pi \right)$ and binning in the appropriate variables.
Since there are typically not enough events to fill the bins with an appropriate sample size, this procedure is carried out tens of times, and the resulting distributions are averaged.

This analysis then uses a binned Poisson likelihood~\cite{ParticleDataGroup:2024cfk} given by:
\begin{equation*}
    \lh(n | \theta) = \prod_{i}\frac{e^{-\mu_{i}\left(\theta\right)}\mu_{i}\left(\theta\right)^{n_{i}}}{n_{i}!}, 
\end{equation*}
where $\theta=\left(\alpha^{\chi},\,\alpha^{\mathrm{sa}},\, \alpha^{\mathrm{a}}\right)$ is the model parameters, i.e. the DM, solar atmospheric, and conventional atmospheric normalizations relative to the nominal; $\mu\left(\theta\right)$ is the expected number of events in a particular model; $n$ is the observed number of events; and $i$ indexes the bin.
To test the hypothesis, the test statistic is given by twice the log-likelihood ratio between the best fit with and without a contribution from DM:
\begin{align*}
    \mathrm{TS} = -2\left[\log\left((n | \lh(0,\, \hat{\alpha}^{\mathrm{sa}},\, \hat{\alpha}^{\mathrm{a}})\right) - \log\left(\lh(n|\check{\alpha}^{\chi},\, \check{\alpha}^{\mathrm{sa}},\, \check{\alpha}^{\mathrm{a}})\right)\right].
\end{align*}
Here, $\hat{\alpha}^{\mathrm{y}}$ and $\check{\alpha}^{\mathrm{y}}$ denote the best-fit normalization in the background-only and signal-plus-background models, respectively.
All background-only test-statistic distributions follow a modified $\chi^{2}$ distribution with one degree of freedom, with the modifications arising from allowing only positive normalizations to be fit, leading to 50\% of events having a test statistic of zero.

\section{Results}
\label{sec:results}

\begin{figure*}[t!]
    \centering
    \begin{subfigure}[t]{0.5\textwidth}
        \centering
        \includegraphics[width=0.95\linewidth]{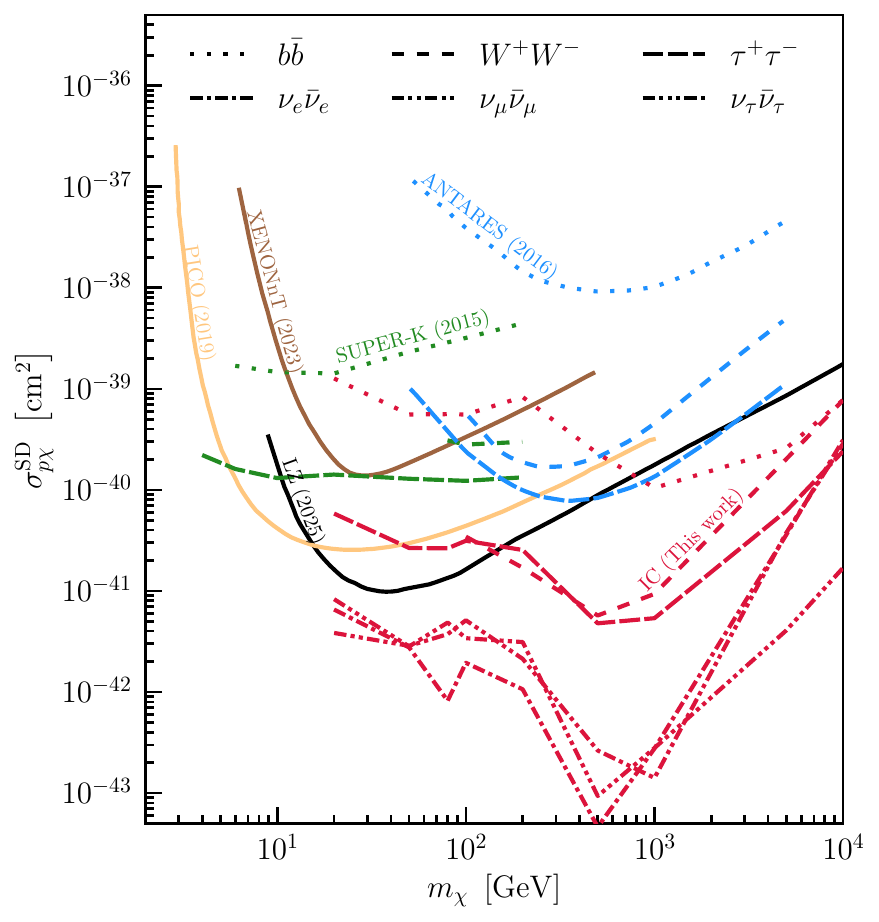}
    \end{subfigure}%
    \hfill
    \begin{subfigure}[t]{0.5\textwidth}
        \centering
        \includegraphics[width=0.95\linewidth]{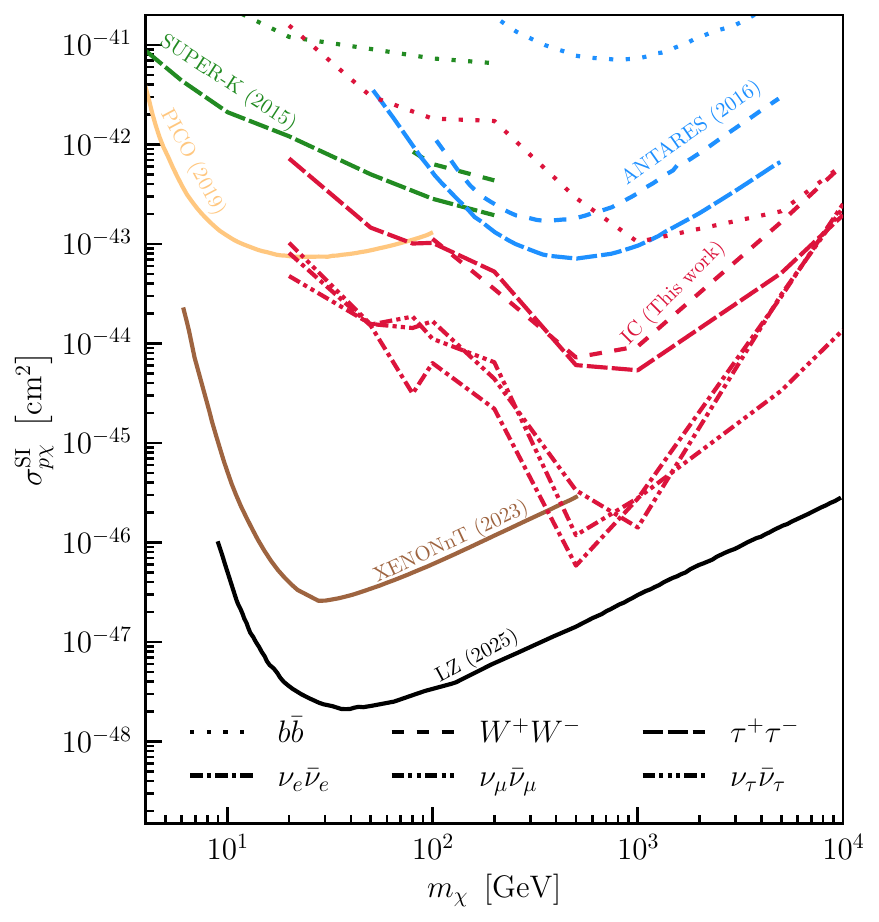}
    \end{subfigure}
    \caption{\textbf{\textit{Limits on the spin-dependent and spin-independent DM-proton cross sections.}}
    Since no excess of neutrinos was observed from the direction of the Sun, limits on the spin-dependent WIMP-proton cross section are obtained.
    These new limits are compared to existing limits on this cross section from other experiments.
    Direct detection experiments are shown with solid lines, whereas indirect detection experiments are shown with line styles corresponding to the annihilation channel.
    This analysis achieves world-leading limits for most annihilation channels for DM above \SI{200}{\GeV}.
    Additionally, this analysis improves on the high-mass $b\bar{b}$ limits by more than an order of magnitude, highlighting the importance of the electroweak correction for hadronic annihilation channels.
    These limits are compared to the previous limits from PICO~\cite{PICO:2019vsc}, ANTARES~\cite{ANTARES:2016xuh}, XENONnT~\cite{XENON:2023cxc}, LUX-ZEPLIN~\cite{LZ:2024zvo}, and Super-Kamiokande~\cite{Super-Kamiokande:2015xms}.
}
    \label{fig:nucleonlimits}
\end{figure*}

Among the 51 dark matter (DM) scenarios examined in this analysis, 32 yielded best-fit results consistent with no DM contribution to the neutrino flux.
Of the remaining 19 scenarios that favored a non-zero DM signal, only the case of 20 GeV DM annihilating into \bb{} showed a preference for the DM-plus-Standard-Model hypothesis with a $p$-value below 0.1---specifically, a $p$-value of 0.079, corresponding to a significance of $1.76\sigma$.
However, due to the low statistical significance and the large number of hypotheses tested, we do not claim evidence for a neutrino excess from DM annihilation in the Sun.

To set upper limits, we vary the DM contribution until the difference in the test statistic from the best-fit point reaches 1.64, corresponding to the 90\% confidence level under the modified $\chi^{2}$ distribution described earlier.
The resulting constraints on spin-dependent and spin-independent DM-proton cross sections are shown in \cref{fig:nucleonlimits}, alongside existing experimental bounds.

\begin{figure}[h]
    \centering
    \includegraphics[width=0.5\linewidth]{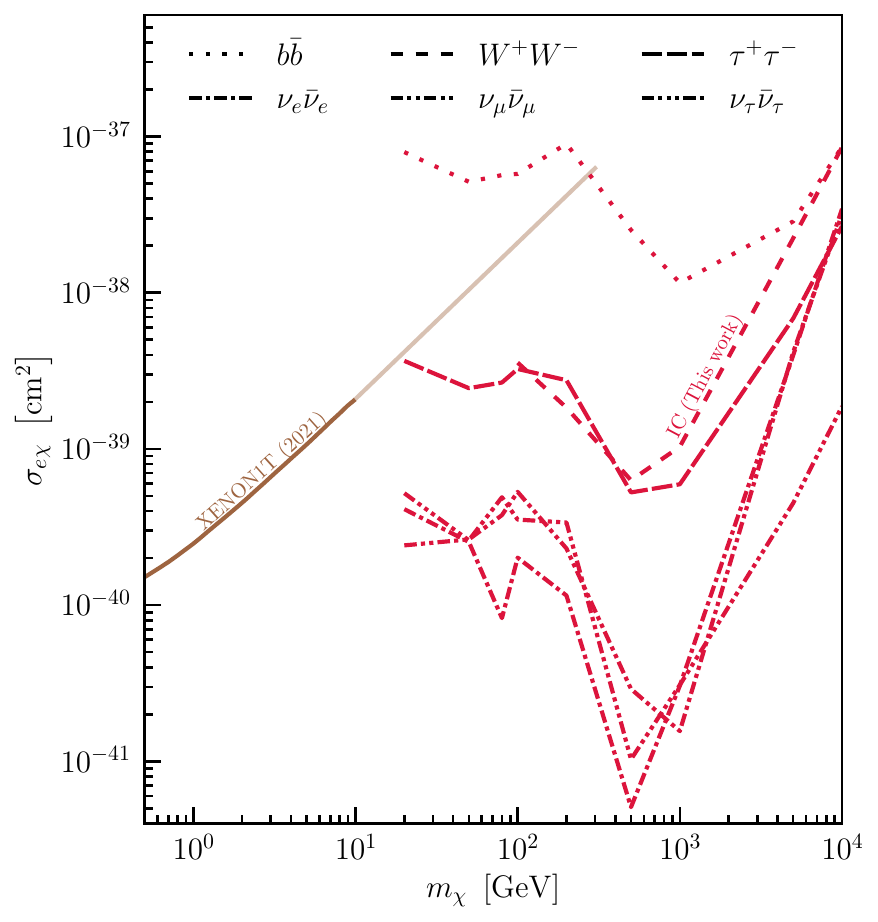}
    \caption{\textbf{\textit{Limits on the DM-nucleon cross section.}}
    These new limits are compared to existing limits on this cross section from other experiments.
    Direct detection experiments are shown with solid lines, whereas indirect detection experiments are shown with line styles corresponding to the annihilation channel.
    These limits are compared to results from the XENON1T~\cite{XENON:2019gfn}.
    Results scaled according to $m_{\chi}$ are shown as a lighter line for comparison at masses higher than had been reported in the original work.
    }
    \label{fig:electron_limits}
\end{figure}

Several noteworthy features emerge from these limits. For DM annihilating to \ww{}, \tautau{}, or \nunu{} pairs, this analysis probes previously unexplored regions of parameter space for DM masses above \SI{200}{\GeV}.
Additionally, the limits on annihilation to \bb{} at high DM masses are significantly improved, enhancing previous IceCube results by roughly a factor of 40, and surpassing ANTARES limits by about a factor of 100.
Overall, this analysis sets the most stringent indirect detection limits for nearly all tested DM scenarios, with the sole exception being the 20 GeV \bb{} case.
These DM-proton cross section limits can also be reinterpreted as constraints on the DM-electron cross section.
The leading global limits obtained from this analysis are presented in \cref{fig:electron_limits}.

\setlength{\bibsep}{0pt plus 0.3ex}
\bibliographystyle{unsrt}  
\bibliography{main} 



%
%
%

\clearpage

\section*{Full Author List: IceCube Collaboration}

\scriptsize
\noindent
R. Abbasi$^{16}$,
M. Ackermann$^{63}$,
J. Adams$^{17}$,
S. K. Agarwalla$^{39,\: {\rm a}}$,
J. A. Aguilar$^{10}$,
M. Ahlers$^{21}$,
J.M. Alameddine$^{22}$,
S. Ali$^{35}$,
N. M. Amin$^{43}$,
K. Andeen$^{41}$
% ...
\\
\\
% ...
$^{10}$ Universit{\'e} Libre de Bruxelles, Science Faculty CP230, B-1050 Brussels, Belgium \\
% ...
$^{16}$ Department of Physics, Loyola University Chicago, Chicago, IL 60660, USA \\
$^{17}$ Dept. of Physics and Astronomy, University of Canterbury, Private Bag 4800, Christchurch, New Zealand \\
% ...
$^{21}$ Niels Bohr Institute, University of Copenhagen, DK-2100 Copenhagen, Denmark \\
$^{22}$ Dept. of Physics, TU Dortmund University, D-44221 Dortmund, Germany \\
% ...
$^{35}$ Dept. of Physics and Astronomy, University of Kansas, Lawrence, KS 66045, USA \\
% ...
$^{39}$ Dept. of Physics and Wisconsin IceCube Particle Astrophysics Center, University of Wisconsin{\textemdash}Madison, Madison, WI 53706, USA \\
% ...
$^{41}$ Department of Physics, Marquette University, Milwaukee, WI 53201, USA \\
% ...
$^{43}$ Bartol Research Institute and Dept. of Physics and Astronomy, University of Delaware, Newark, DE 19716, USA \\
% ...
$^{63}$ Deutsches Elektronen-Synchrotron DESY, Platanenallee 6, D-15738 Zeuthen, Germany \\
$^{\rm a}$ also at Institute of Physics, Sachivalaya Marg, Sainik School Post, Bhubaneswar 751005, India \\
% ...

\subsection*{Acknowledgments}

\noindent
The authors gratefully acknowledge the support from the following agencies and institutions:
USA {\textendash} U.S. National Science Foundation-Office of Polar Programs,
U.S. National Science Foundation-Physics Division,
U.S. National Science Foundation-EPSCoR,
U.S. National Science Foundation-Office of Advanced Cyberinfrastructure,
Wisconsin Alumni Research Foundation,
Center for High Throughput Computing (CHTC) at the University of Wisconsin{\textendash}Madison,
Open Science Grid (OSG),
Partnership to Advance Throughput Computing (PATh),
Advanced Cyberinfrastructure Coordination Ecosystem: Services {\&} Support (ACCESS),
Frontera and Ranch computing project at the Texas Advanced Computing Center,
U.S. Department of Energy-National Energy Research Scientific Computing Center,
Particle astrophysics research computing center at the University of Maryland,
Institute for Cyber-Enabled Research at Michigan State University,
Astroparticle physics computational facility at Marquette University,
NVIDIA Corporation,
and Google Cloud Platform;
Belgium {\textendash} Funds for Scientific Research (FRS-FNRS and FWO),
FWO Odysseus and Big Science programmes,
and Belgian Federal Science Policy Office (Belspo);
Germany {\textendash} Bundesministerium f{\"u}r Forschung, Technologie und Raumfahrt (BMFTR),
Deutsche Forschungsgemeinschaft (DFG),
Helmholtz Alliance for Astroparticle Physics (HAP),
Initiative and Networking Fund of the Helmholtz Association,
Deutsches Elektronen Synchrotron (DESY),
and High Performance Computing cluster of the RWTH Aachen;
Sweden {\textendash} Swedish Research Council,
Swedish Polar Research Secretariat,
Swedish National Infrastructure for Computing (SNIC),
and Knut and Alice Wallenberg Foundation;
European Union {\textendash} EGI Advanced Computing for research;
Australia {\textendash} Australian Research Council;
Canada {\textendash} Natural Sciences and Engineering Research Council of Canada,
Calcul Qu{\'e}bec, Compute Ontario, Canada Foundation for Innovation, WestGrid, and Digital Research Alliance of Canada;
Denmark {\textendash} Villum Fonden, Carlsberg Foundation, and European Commission;
New Zealand {\textendash} Marsden Fund;
Japan {\textendash} Japan Society for Promotion of Science (JSPS)
and Institute for Global Prominent Research (IGPR) of Chiba University;
Korea {\textendash} National Research Foundation of Korea (NRF);
Switzerland {\textendash} Swiss National Science Foundation (SNSF).

\end{document}